\documentclass{article}
\usepackage{graphicx} 
\usepackage[english]{babel}
\usepackage[a4paper, left=1cm]{geometry}
\title{ANALYTICAL MODEL OF PHOTON RING WITH FINITE THICKNESS}

\author{S.~V. Chernov \\
Astro Space Center of P. N. Lebedev Physical Institute, Moscow, Russia\\
email: chernov@td.lpi.ru}
\date{}
\begin{document}

\maketitle

\abstract{An analytical model of a thick asymmetric Gaussian ring is presented for which the visibility function is calculated in two perpendicular directions for baselines up to 6 of the Earth’s diameter.}

\section{Introduction}

Recently, the Event Horizon Telescope (EHT)
team published the first images of supermassive black
holes at the centers of two galaxies: the elliptical
galaxy M$87^*$ [1] and our Milky Way galaxy, Sgr $A^*$
[2]. The observations were made in the millimeter
band at 230 GHz. The images obtained clearly show
an asymmetric inhomogeneous bright ring with a
dark spot in its center. This image is interpreted as
an image of a supermassive black hole [1, 2]. The
diameter of the ring is 42 microseconds of arc ($\mu as$)
for M$87^*$ [1] and 51.8 $\mu as$ for Sgr $A^*$ [2],
respectively. However, due to insufficient angular
resolution (on the order of 25 $\mu as$) and poor
image quality, the EHT group was unable to estimate
the thickness and analyze the structure of the ring in
detail.

The general theory of relativity predicts that
black holes can manifest themselves as a shadow
surrounded by an infinite number of nested photon
rings [3]. Each photon ring corresponds to a certain
number of half-orbits that a photon travels through
before reaching the observer. The photon ring appears
as a bright and distinct feature in the image of the
black hole. It is believed that in the observed images
of black holes obtained by the EHT group, the bright
ring must contain an infinite number of photon rings.
The brightest and widest photon ring is the first, and
it is assumed that the flux and width of subsequent
rings decreases by $e^{-\pi}$ factor [4, 5]. Measuring and
analyzing the shape of the photon ring will allow the
spin of the black hole to be estimated.

To date, concepts of space radio telescopes that
can observe the first photon ring have been considered
[6, 7]. In [6], space radio telescopes in circular regular
and retrograde orbits are discussed. This concept
allows high quality images of supermassive black
holes with sufficiently large angular resolution down
to $\sim 1-2$ $\mu as$. Using this concept, it is possible
to resolve the ring obtained by the EHT group and
observe the first photon ring of a supermassive black
hole. As we know, an infinitely thin ring will manifest
at bases larger than the Earth’s diameter as universal
fast oscillating functions on the visibility function
graph [5, 8, 9]. A thick ring, on the contrary, will
manifest at bases smaller than the Earth’s diameter.
Modeling numerical observations and analyzing the
prominence functions at bases larger than the Earth’s
diameter will help to better understand and study the
nature of black hole physics.

In this paper, we consider an analytical model
of a thick asymmetric ring with a radial Gaussian
brightness profile and calculate the visibility function
with bases up to 6 Earth diameters at 690 GHz.

\section{ANALYTICAL MODEL AND RESULTS}

The response of a radio interferometer to the
signal distribution on the sky is a complex visibility
function. This function is related to the brightness
distribution of the source on the celestial sphere
$I(r,\phi_r)$ through the Fourier transform [10]. In polar
coordinates $u,\phi_u$ , the visibility function is defined as
follows [10]
\begin{eqnarray}
 V(u,\phi_u)=\int\int I(r,\phi_r) e^{-2\pi iur\cos(\phi_r-\phi_u)}rdrd\phi_r,
 \label{vid}
\end{eqnarray}
where $u$ is the dimensionless projection of the
interferometer base, expressed in units of wavelengths,
and $r,\phi_r$ are the polar coordinates in the picture
plane of the source, expressed in radians.

Here we will assume that the brightness
distribution function of the source on the sky can be
represented as the product of the radial and azimuthal
components:
\begin{eqnarray}
 I(r,\phi_r)=I_r(r)I_\phi(\phi_r).
 \label{brightness}
\end{eqnarray}
We will describe the radial component by a Gaussian
distribution function as follows:
\begin{eqnarray}
 I_r(r)=e^{-\left(\frac{r-r_0}{\Delta r}\right)^2},
 \label{rbrightness}
\end{eqnarray}
where $r_0$ is the radius and $\Delta r$ is the width of the
ring in radians. The azimuthal part is an asymmetric
distribution function given in the form of
\begin{eqnarray}
 I_\phi(\phi_r)=\left(1-B\sin^2\frac{\phi_r-\phi_0}{2}\right)^n,
 \label{axibrightness}
\end{eqnarray}
where B is the asymmetry parameter, $\phi_0$ is the
direction to the maximum brightness in the ring,
and n is the degree of asymmetry. If the asymmetry
parameter is zero $B = 0$ , we obtain a symmetric
distribution function. In this case, the visibility
function is calculated quite simply (see [8]). It
should be noted that in more realistic scenarios
the asymmetry parameter B can depend on the
spin of the black hole and the inclination angle.
In this paper, we will consider the case where
the asymmetry parameter is a constant quantity
independent of spin and inclination angle. In
addition, we will consider the case when the degree
of asymmetry is equal to one, $n = 1$.

Figure 1 shows an example of the brightness
distribution in the ring for a radius $r_0 = 20$ $\mu as$ and
ring width $\Delta r = 5$ $\mu as$. The brightness maximum is
located at the angle $\phi_0=\pi$. The total flux in the ring
is 1 Jansky. If we substitute the expression (4) into
the equation (1), the integral over the angle fr can
be taken explicitly. As a result, we obtain
\begin{eqnarray}
 V(u,\phi_u)=\pi\int [(2-B)J_0(2\pi ur)
 -iBJ_1(2\pi ur)\cos(\phi_u-\phi_0)]I_r(r)rdr,
 \label{V1}
\end{eqnarray}
where $J_0$ and $J_1$ are the Bessel functions of the
first kind of zero and first order, respectively. This
expression (5) is general, and the visibility function
is a complex quantity due to the asymmetry of
the sky brightness distribution function. In the
symmetric case, when $B = 0$, the visibility function
becomes a real quantity. It is also worth noting that
the visibility function depends on the direction of
projection of the interferometer base, i.e., on the
angle $\phi_u$. In different directions, the amplitude and
phase of the visibility function can take different
values.

The integral (5) with the distribution function
(3) can be calculated with good accuracy by
approximate methods. We will be primarily interested
in sufficiently large bases larger than the diameter of
the Earth. For such bases, the Bessel function can
be expanded into a series over a large argument
($2\pi u r\gg1$, see [11])
\begin{eqnarray}
J_0(x)\approx\frac{\sin(x)+\cos(x)}{\sqrt{\pi x}},\\
J_1(x)\approx\frac{\sin(x)-\cos(x)}{\sqrt{\pi x}}.
\label{bessel}
\end{eqnarray}
It is worth noting that this decomposition is also valid
for bases smaller than the Earth’s diameter. Then the
visibility function can be rewritten as

\begin{eqnarray}
V(u,\phi_u)=\frac{2-B}{\sqrt{2u}}\int_0^\infty(\sin(2\pi ur)+\cos(2\pi ur))\sqrt{r}e^{-(\frac{r-r_0}{\Delta r})^2}dr-\nonumber\\
 -iB\frac{\cos(\phi_u-\phi_0)}{\sqrt{2u}}\int_0^\infty(\sin(2\pi ur)-\cos(2\pi ur))\sqrt{r}e^{-(\frac{r-r_0}{\Delta r})^2}dr.
 \label{V2}
\end{eqnarray}

\begin{figure}
\centering
\includegraphics[width=1.1\linewidth]{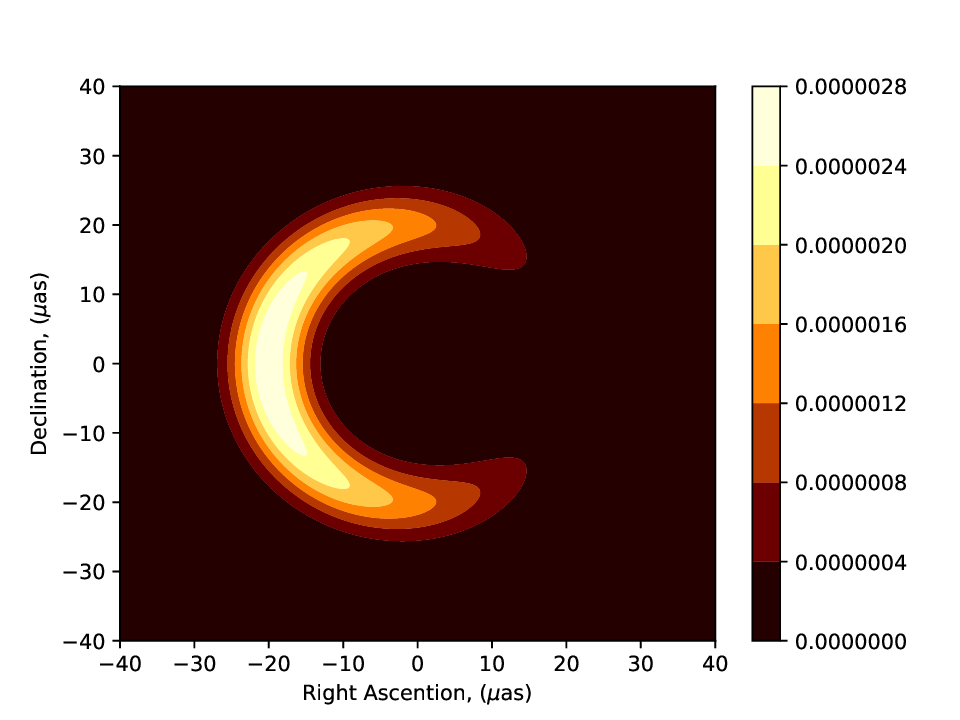}
\caption{Brightness distribution in the ring for the following parameters: $r_0=20 \mu as$, $\Delta r=5 \mu as$, $\phi_0=\pi$, $B=1$ and $n=1$. The total flux in the ring is equal to 1 Jansky.}
\label{fig1}
\end{figure}

This integral (8) can be calculated analytically exactly using the expression (3.953) from [12]. As a result, we obtain a final expression for the visibility function with an asymmetric brightness distribution (4) and a radial Gaussian profile (3)
\begin{eqnarray}
V(u,\phi_u)=\frac{2-B}{\sqrt{8u}}\Gamma\left(\frac{3}{2}\right)\left(\frac{\Delta r^2}{2}\right)^{3/4}\exp\left(-\frac{r_0^2}{2\Delta r^2}-\frac{\pi^2 u^2\Delta r^2}{2}\right)\times\nonumber\\
 \times\bigg[(1-i)\exp\left(i\pi ur_0\right)D_{-3/2}\left(-i\sqrt{2}\pi u\Delta r-\sqrt{2}\frac{r_0}{\Delta r}\right)
 +(1+i)\exp\left(-i\pi ur_0\right)D_{-3/2}\left(i\sqrt{2}\pi u\Delta r-\sqrt{2}\frac{r_0}{\Delta r}\right)\bigg]-\nonumber\\
 -iB\frac{\cos(\phi_u-\phi_0)}{\sqrt{8u}}\Gamma\left(\frac{3}{2}\right)\left(\frac{\Delta r^2}{2}\right)^{3/4}\exp\left(-\frac{r_0^2}{2\Delta r^2}-\frac{\pi^2 u^2\Delta r^2}{2}\right)\times\nonumber\\
 \times\bigg[-(1+i)\exp\left(i\pi ur_0\right)D_{-3/2}\left(-i\sqrt{2}\pi u\Delta r-\sqrt{2}\frac{r_0}{\Delta r}\right)
 +(i-1)\exp\left(-i\pi ur_0\right)D_{-3/2}\left(i\sqrt{2}\pi u\Delta r-\sqrt{2}\frac{r_0}{\Delta r}\right)\bigg]
 \label{V-Weber}
\end{eqnarray}
where D is the Weber function (parabolic cylinder function). The visibility function (9) is a complex function
and depends on both the radius and width of the ring as well as the base direction $\phi_u$. For convenience in
computing the visibility function and visualization, we can rewrite the Weber functions through Hermite
polynomials. As a result, we obtain
\begin{eqnarray}
V(u,\phi_u)=\frac{2-B}{\sqrt{8u}}\Gamma\left(\frac{3}{2}\right)\left(\Delta r\right)^{3/2}(1-i)\exp(-\frac{r_0^2}{\Delta r^2})
\bigg[H_{-3/2}\left(-i\pi u\Delta r-\frac{r_0}{\Delta r}\right)+iH_{-3/2}\left(i\pi u\Delta r-\frac{r_0}{\Delta r}\right)\bigg]-\nonumber\\
-iB\frac{\cos(\phi_u-\phi_0)}{\sqrt{8u}}\Gamma\left(\frac{3}{2}\right)\left(\Delta r\right)^{3/2}(1+i)\exp(-\frac{r_0^2}{\Delta r^2})
\bigg[iH_{-3/2}\left(i\pi u\Delta r-\frac{r_0}{\Delta r}\right)-H_{-3/2}\left(-i\pi u\Delta r-\frac{r_0}{\Delta r}\right)\bigg].
\label{V-Her}
\end{eqnarray}
where $H$ are Hermite polynomials. The normalization of the visibility function is determined by the expression of the visibility function at small bases, which takes the following form
\begin{eqnarray}
V(u\rightarrow0,\phi_u)\approx
\frac{2-B+iB\cos(\phi_u-\phi_0)}{\sqrt{2u}}\Gamma\left(\frac{3}{2}\right)\left(\Delta r\right)^{3/2}\exp(-\frac{r_0^2}{\Delta r^2})H_{-3/2}\left(-\frac{r_0}{\Delta r}\right).
\end{eqnarray}

It should be noted that at small base projections,
when $u\rightarrow0$, the approximation (7) is no longer
satisfied. This leads to an insignificant difference in
amplitude, which is easily compensated by the fitting
multiplier.

Fig. 2 shows the visibility function in two
perpendicular directions: $u=0,\pi/2$. The
dependence is shown on the base projection expressed
in wavelengths (lower axis) and in Earth diameters
(upper axis). The green and red curves correspond to
the results of numerical integration of the prominence
function (1) for the source (2). The integration was
performed using the finufft package version 1.1.1 (see
[13]). The green curve corresponds to the direction of
$\phi_u=0$, and the red curve corresponds to $\phi_u=\pi/2$.
It is important to note that at $\phi_u=\pi/2$ the visibility
function is real, whereas at $\phi_u=0$ it becomes
complex. This difference is due to asymmetry in the
luminance distribution, which affects the amplitude
and phase of the visibility function. The black curves
represent the analytical solutions obtained from the
formula (10). In general, as it is clear from Fig. 2,
the agreement between the analytical and numerical
curves is quite good.

\begin{figure}
\centering
\includegraphics[width=1.25\linewidth]{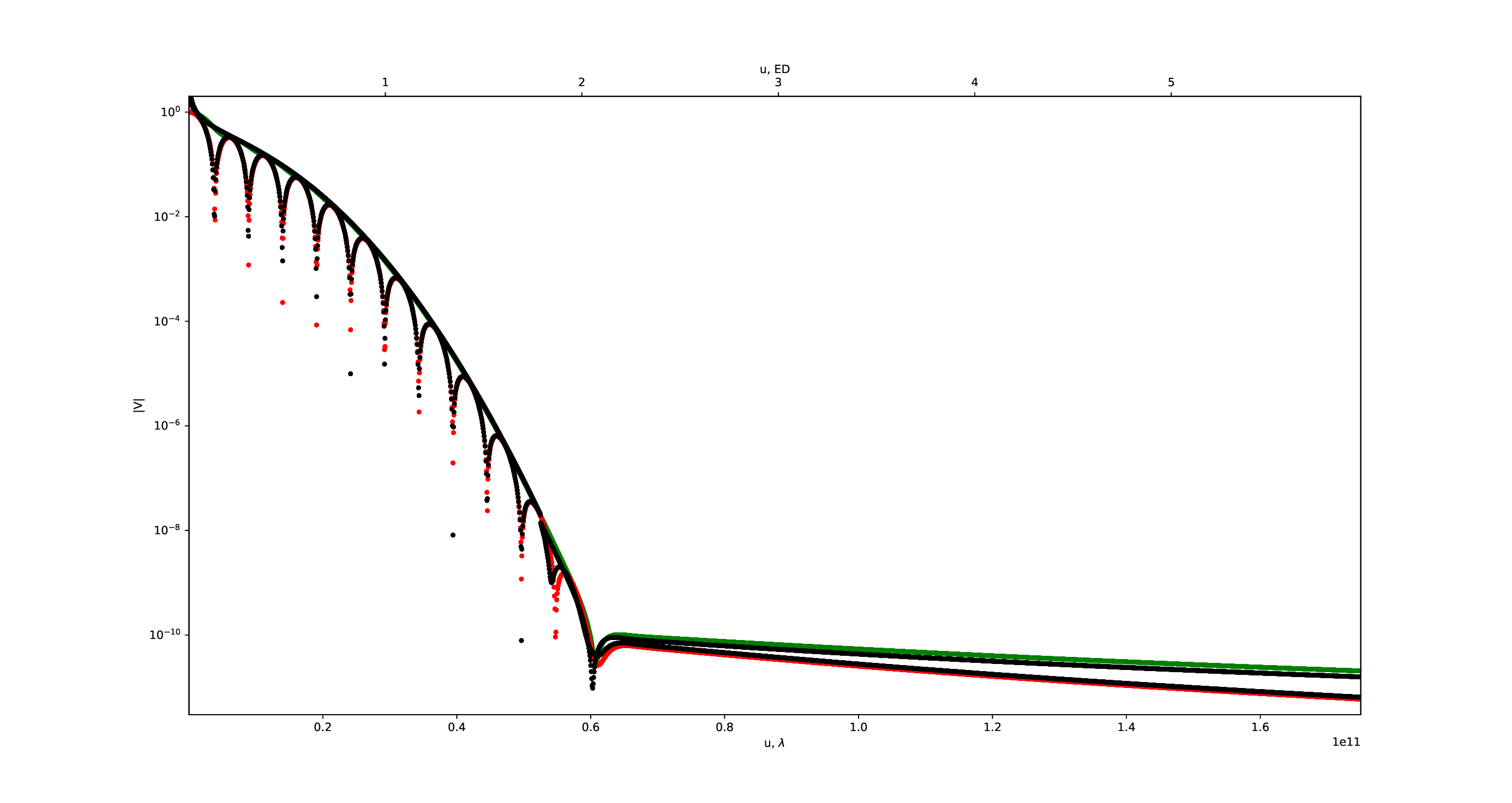}
\caption{The visibility function as a function of the base projection
u is plotted, where the lower axis represents values in wavelength
units and the upper axis in Earth diameters. The green and red
curves are obtained by numerical integration of the expression
(1). The green curve corresponds to the direction of $\phi_u=0$, the
red curve corresponds to the direction of $\phi_u=\pi/2$. Black curves
are analytical curves obtained by the formula (10).}
\label{fig2}
\end{figure}

\section{Conclusion}

In this paper we have constructed an analytical model
of a thick asymmetric ring with a radial Gaussian profile
of brightness distribution. The visibility function was
analytically calculated for this model. It was shown that
the visibility function is a complex function and depends
on both the radius and width of the ring. In addition,
the visibility function depends on the direction of the
base, i.e., the angle fu. The amplitudes of the visibility
function in two perpendicular directions were also
calculated both numerically and analytically using the
formula (10)). It was shown that the analytical formulas
(10) agree quite well with the numerical calculations,
which confirms the correctness of the proposed model
and analysis methods. This model can be directly used
to fit observational data in various missions [7]. Thus,
the obtained results can be useful for further studies in
the field of radio interferometry with ultra-long bases,
as well as for practical applications in observations of
images of supermassive black holes.

\section{REFERENCES}

1.	 The Event Horizon Telescope Collaboration
Astrophys. J. Lett. 875, L1 (2019).

2.	 The Event Horizon Telescope Collaboration
Astrophys. J. Lett. 930, L12 (2022).

3.	 Luminet J.P. A$\&$A 75, 228 (1979).

4.	 Tiede P., Johnson M.D., Pesce D.W., Palumbo D.C.M.,
Chang D.O., Galison P. Galaxies 10, 111 (2022).

5.	 Johnson M.D., et al. Sci. Adv. 6, 1310 (2020).

6.	 Rudnitskiy A.G., Shchurov M.A., Chernov S.V.,
Syachina T.A., Zapevalin P.R. Acta Astronaut. 212,
361 (2023).

7.	 Likhachev S.F., Rudnitskiy A.G., Shchurov M.A.,
Andrianov A.S., Baryshev A.M., Chernov S.V.,
Kostenko V.I. MNRAS511, 668 (2022).

8. Chernov S.V. JETP 159, 1018 (2021).

9. Chernov S.V., Shchurov M.A ., Bulygin I.I.,
Rudnitskiy A.G. arXiv:2502.03026.

10.	 Thompson A .R., Moran J.M., Swenson G.W.
Interferometry and Synthesis in Radio Astronomy,
2nd ed. (Moscow: Fizmatlit, 2003).

11.	 Zaitsev V.F., Polyanin A.D. Handbook of Ordinary
Differential Equations (Moscow: Fizmatlit, 2001).

12.	 Gradshteyn I.S., Ryzhik I.M. Tables of Integrals, Sums,
Series, and Products (Moscow: State Publishing
House of Physical and Mathematical Literature,
1963).

13.	 Barnett A.H., Magland J.F., af Klinteberg L. SIAM
J. Sci. Comput. 41, 479 (2019)

\end{document}